# Impact of Dispersive and Saturable Gain/Loss on Bistability of Nonlinear Parity-Time Bragg Gratings

Sendy Phang,[1,*] Ana Vukovic,[1] Hadi Susanto,[2] Trevor M. Benson,[1] and Phillip Sewell[1]

[1]*The George Green Institute for Electromagnetics Research, Faculty of Engineering, University of Nottingham, Nottingham, NG7 2RD, UK*
[2]*Department of Mathematical Sciences, University of Essex, Colchester CO4 3SQ, UK*
*\*Corresponding author: sendy.phang@nottingham.ac.uk*

We report on the impact of realistic gain and loss models on the bistable operation of nonlinear parity-time Bragg gratings. In our model we include both dispersive and saturable gain and show that levels of gain/loss saturation can have significant impact on the bistable operation of a nonlinear PT Bragg grating based on GaAs material. The hysteresis of the nonlinear PT Bragg grating is analyzed for different levels of gain and loss and different saturation levels. We show that high saturation levels can improve the nonlinear operation by reducing the intensity at which the bistability occurs. However when the saturation intensity is low, saturation inhibits the PT characteristics of the grating.

Optical structures with balanced gain and loss, mimicking parity and time (PT) symmetry in quantum field theory [1], have been the subject of intense investigation in the last few years. PT symmetric structures based on Bragg gratings [2–4], couplers [5,6], and lattices [7,8], have been reported and demonstrated functionalities including optical switching [4,9–11], unidirectional invisibility [2,7,12], and memory [13]. Unidirectional invisibility [7,12] and power oscillation [14] have also been experimentally demonstrated. A linear PT-symmetric Bragg grating (PTBG) has a different response for a signal incident from the left and right side of the grating whereby the transmittances are the same, $T_L = T_R$, and the reflectances are different, $\Gamma_L \neq \Gamma_R$. It is important to note that although this is commonly referred to as a non-reciprocal behavior, in a strict sense a linear PT structure does satisfy the Lorentz reciprocity condition [15,16]. This is because the scattering matrix of the structure is complex-symmetric $\bar{\bar{S}} = (\bar{\bar{S}})^t$, although it is no longer unitary or orthogonal $\bar{\bar{S}} = (\bar{\bar{S}})^\dagger$, where $t$ and $\dagger$ represent the transpose and transpose-conjugate operators, respectively. Of particular interest is the unidirectional invisibility phenomenon, which occurs when the modulation of real and imaginary part of the refractive index of the structure are equal, at which no reflection is observed from one side of the grating. In the case of linear and frequency independent materials unidirectional invisibility is present at all frequencies [2–4]. It has been suggested that the inclusion of Kerr-type nonlinearity into the PT gratings promises to open a range of new applications, or to improve the existing ones [2,9,13]. It is important to note that a few papers that considered nonlinear PT structures [2,9,17] have done so under the assumptions that the gain and loss are non-saturable and non-dispersive.

In this paper we extend the analysis by considering a nonlinear PT Bragg grating that has both dispersive and saturable gain and loss with the real and imaginary parts of refractive index satisfying the Kramers-Kronigs relationship. In particular, we analyze the unidirectional operation of dispersive Bragg gratings and then extend the analysis to nonlinear PT Bragg gratings and report on how different levels of gain and loss saturation can have a crucial role in enabling or inhibiting the interplay between the PT and nonlinear behavior. For this we consider a scenario of a GaAs Bragg grating with realistic parameters of material dispersion, nonlinearity and gain/loss saturation. For this purpose a time-stepping numerical technique based on the Transmission Line modeling (TLM) method [18] is used. The TLM method is a flexible time-stepping numerical technique that has been extensively characterized and used over many years [19]. However any time-domain method, including the finite-difference time-domain (FDTD) method, could be employed for this purpose. It is appropriate to comment that the TLM method has comparable performance with the FDTD method but offers certain advantages for particular applications [19]. In our earlier work we have validated the TLM method for modeling PT Bragg gratings, and used it to demonstrate real-time optical switching assuming a simple case where material and gain/loss models are frequency and intensity independent [4].

A nonlinear PT Bragg grating (NPTBG) is illustrated in Fig. 1(a). The structure is embedded in a background material with a refractive index $n_B$ and has a length of $N\Lambda$, where $\Lambda$ denotes the length of a single period and $N$ is the total number of periods. The refractive index distribution in a single period, $\hat{n}_G$, along the propagation direction $z$, shown in Fig. 1(b) can be expressed as,

$$\hat{n}_G(z,\omega,I,t) = \begin{cases} \hat{n}_H(\omega) + n_2 I(z,t) + j\frac{c}{\omega}\alpha(\omega,I), & z < \frac{\Lambda}{4} \\ \hat{n}_L(\omega) + n_2 I(z,t) + j\frac{c}{\omega}\alpha(\omega,I), & \frac{\Lambda}{4} < z < \frac{\Lambda}{2} \\ \hat{n}_L(\omega) + n_2 I(z,t) - j\frac{c}{\omega}\alpha(\omega,I), & \frac{\Lambda}{2} < z < \frac{3\Lambda}{4} \\ \hat{n}_H(\omega) + n_2 I(z,t) - j\frac{c}{\omega}\alpha(\omega,I), & \frac{3\Lambda}{4} < z < \Lambda \end{cases} \quad (1)$$

where $\hat{n}_H$ and $\hat{n}_L$ are the complex high and low refractive indices which are frequency dependent, $n_2$ is the Kerr nonlinearity constant, $I$ is the input signal intensity and $\pm\alpha(\omega,I)$ denotes the gain(+) or loss(-) in the grating that is both dispersive and saturable. The complex dispersive dielectric material based on a simple harmonic oscillator model with a Lorentzian profile is implemented in the TLM method, in which the refractive index $\hat{n}$ at any frequency $\omega$ can be calculated as,

$$\hat{n}^2 = (1 + \chi_{e\infty}) + \frac{\chi_{e0}\omega_{0D}^2}{2j\omega\delta + (\omega_{0D}^2 - \omega^2)} \quad (2)$$

Here, $\delta$ and $\omega_{0D}$ denote the damping and the resonant frequency of the medium, $\chi_{e\infty}$ denote the dielectric susceptibility at infinite frequency, while $\chi_{e0}$ is the dispersive dielectric susceptibility contribution to the overall material refractive index. Detail of the implementation and validation of the material dispersion and nonlinearity in the TLM method has been reported in [20]. The Bragg frequency $f_B$ of the grating is related to the real part of average refractive index $\bar{n} = \frac{1}{2}\text{Re}(\hat{n}_H+\hat{n}_L)$ of the structure by $f_B = \frac{c}{2\bar{n}\Lambda}$, where $c$ is the phase velocity of light in free-space.

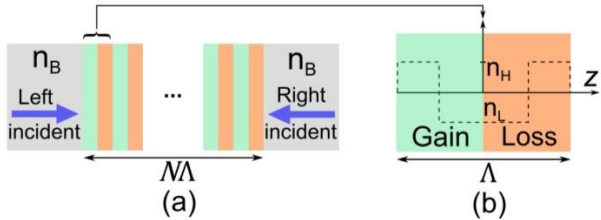

Fig. 1 (a) Schematic of a $N$-period PTBG in a background material $n_B$, (b) single period of a grating with $n_H$ and $n_L$ as the high and low real refractive index at $f_B$ in even symmetry and gain (green) and loss (orange) in odd symmetry

A saturable and dispersive gain/loss model exhibiting homogenous broadening with a Lorentzian profile is implemented in the TLM method as [21],

$$|\alpha|(\omega,I) = \mathbb{S}(I)\left(\frac{|\alpha_0|}{1+j(\omega-\omega_{0\sigma})\tau} + \frac{|\alpha_0|}{1+j(\omega+\omega_{0\sigma})\tau}\right) \quad (3)$$

Here, the gain/loss per unit length $|\alpha|$ is related to the imaginary part of refractive index as $|\alpha| = \frac{\omega}{c}|n_I|$, $\omega_{0\sigma}$ denotes the atomic-transitional angular-frequency, $\tau$ is the dipole-relaxation time parameter and $\alpha_0$ denotes the peak value of gain or loss per unit length at $\omega_{0\sigma}$. The intensity dependent function $\mathbb{S}(I)$ describes the saturation factor of the gain or loss as,

$$\mathbb{S}(I) = \frac{1}{1 + I/I_S} \quad (4)$$

where $I$ and $I_S$ denote the input signal and saturation intensity, respectively. The saturation factor $\mathbb{S}$ varies between $0 < \mathbb{S} < 1$, with $\mathbb{S} = 0$ denoting a high saturation level ($\frac{I}{I_S} \to \infty$), whilst $\mathbb{S} = 1$ denotes the unsaturated state ($\frac{I}{I_S} \to 0$). It is important to note that the model (2)-(3) satisfies the Kramers-Kronigs relationship between the real and imaginary part of refractive index of material.

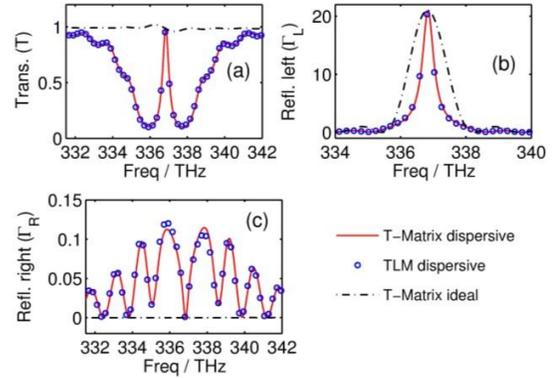

Fig. 2.(a) Transmittance, (b) reflectance for left and (c) right incident of PTBG as a function of frequency considering a linear medium ($n_2 = 0$) at U condition ($|\alpha_0| = 1460.24$ cm$^{-1}$) with no gain/loss saturation ($\mathbb{S} = 1$). The results obtained using the T-matrix method for the idealized gain/loss model are included for references.

Throughout this paper 200 periods of a PTBG based on GaAs material are considered with the following material parameters, $\chi_{e0} = 7.5$, $\omega_{0D} = 4614.4$ rad/ps, and $\delta = 0.0923$ rad/ps [22], with the high and low refractive index i.e. $n_H$ and $n_L$ obtained at the Bragg frequency from the high and low dielectric susceptibilities, $\chi_{e\infty} = 2.8$ and 2.5 respectively, which form the grating. It is here noted that a small change in the refractive index can be achieved by replacing a small amount of Ga by Al as [23]. For modelling purposes this small change in refractive index is represented by a different $\chi_{e\infty}$. The Kerr nonlinear constant is $n_2 = 2 \times 10^{-17}$ m$^2$W$^{-1}$ [24] throughout the structure. The gain and loss parameters are $\tau = 0.1$ ps and $\omega_{0\sigma} = 2116.5$ rad/ps [21], while $\alpha_0$ depends on the gain and loss given. The background material refractive index of GaAs at $\omega_{0\sigma}$ is $n_B = 3.626$. The periodicity of the NPTBG is designed so that the band-gap of the structure is centered at the atomic-transitional frequency, i.e. $f_B = \frac{\omega_{0\sigma}}{2\pi}$, hence $\Lambda = 122.7$ nm. The unidirectional (U) operation of the PTBG happens when the gain/loss parameter $|\alpha_0| = \frac{1}{2}\frac{\omega}{c}(n_H - n_L)$, [4]. For the chosen material parameters, we obtain unidirectional operation at $f_B$ when the gain and loss coefficient $|\alpha_0| = 1460.24$ cm$^{-1}$. All simulations were done using the TLM method and for good accuracy a spatial discretization of $\Delta z = c/(96 f_B)$ as reported in [4]. The frequency domain response is obtained by Fourier transformation of the time-domain signal. The TLM simulation was run for $NT = 524288$ time-steps which ensured that the entire signal has passed through the structure and provided a sufficient frequency-domain resolution.

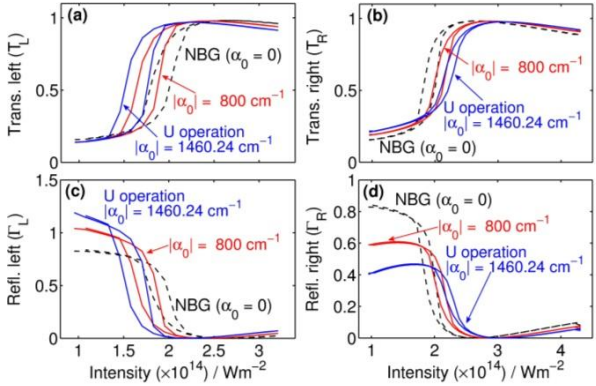

Fig. 3. Hysteresis of a NPTBG with high saturation intensity and for different gain/loss parameter $|\alpha_0| = 800$ cm$^{-1}$ and 1460.24 cm$^{-1}$ (U operation); (a) transmittance $T_L$, (c) reflectance $\Gamma_L$, for the signal incident from the left, (b) transmittance $T_R$, (d) reflectance $\Gamma_R$ for the signal incident from the right of the grating. Saturation intensity is $I_S = 5 \times 10^{13}$ Wm$^{-2}$. Dashed line represents the response of the NBG for reference.

Fig. 2 analyzes the impact of material dispersion on the frequency domain response of a linear PTBG ($n_2 = 0$) at the U condition ($|\alpha_0| = 1460.24$ cm$^1$) with no gain/loss saturation ($\mathbb{S} = 1$). The results of the Transfer-matrix (T-matrix) method with and without the dispersive material model are compared with ones calculated by the TLM method. The methodology of the T-matrix method is not described in this paper and the reader is referred to [25]. Fig. 2 shows that results calculated by the TLM agree well with the ones calculated by the T-matrix method. As expected, transmittance is the same for the left and right incidence while the reflectances differ. More importantly, the results obtained with the dispersive material model differ significantly from the ones with non-dispersive material parameters in that unidirectional invisibility ($T \rightarrow 1, \Gamma_R \rightarrow 0$) is not observed at all frequencies but is confined to a narrowband region centered at the Bragg frequency. We believe that this is an important result that limits the PT structures in real cloaking applications.

We now consider the case of a nonlinear PTBG (NPTBG) and analyze how different levels of gain/loss saturation affect the performance of the grating. Fig. 3 shows the response of the NPTBG with a high gain/loss saturation intensity of $I_S = 5 \times 10^{13}$ Wm$^{-2}$. Fig. 3 shows (a) transmittance $T_L$ and (c) reflectance $\Gamma_L$ for the left incident and (b) transmittance $T_R$ and (d) reflectance $\Gamma_R$ for the right incident signal as a function of input signal intensity and for different gain and loss parameter $|\alpha_0|$. It is noted that for the given variation of input signal intensity, the saturation factor $\mathbb{S}$ varies between $0.3 < \mathbb{S} < 0.6$ although it is emphasized that saturation factor throughout the structure varies due to diffraction and the presence of gain and loss. For comparison, the response of a nonlinear Bragg grating (NBG) (i.e. one without gain and loss, $|\alpha_0| = 0$) is depicted by dashed lines. In order to obtain bistable operation the input signal frequency is set to be at the band-gap edge [26], in which we pick a continuous-wave (CW) operating at $f = 337.7$ THz. The hysteresis is obtained by gradually increasing and decreasing the input signal intensity in a single computation. This is repeated for different gain/loss parameters, that is $|\alpha_0| = 800$ cm$^{-1}$ and 1460.24 cm$^{-1}$ (U operation). It is here emphasized that U operation denotes the grating parameters and not the resulting grating response.

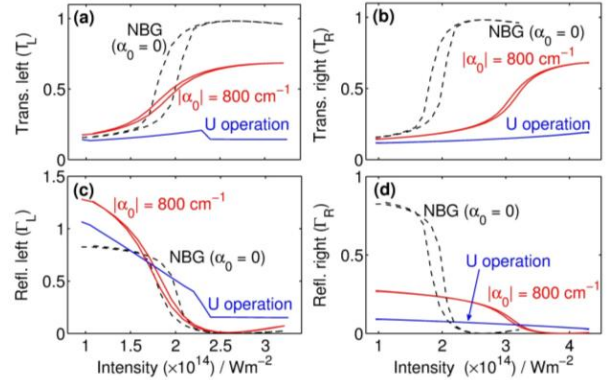

Fig. 4. Hysteresis of NPTBG operating at $f = 337.7$ THz as a function of input signal intensity for two different values of gain/loss parameters, $|\alpha_0| = 800$ cm$^{-1}$ and 1460.24 cm$^{-1}$ (U operation). Gain and loss saturation is turned off ($\mathbb{S} = 1$). (a) transmittance $T_L$ and (c) reflectance $\Gamma_L$ for left incidence, (b) the transmittance $T_R$ and (d) reflectance $\Gamma_R$ for the right incidence. Hysteresis of the NBG is included for reference.

Fig. 3(a-d) shows that the NPTBG is bistable for both transmittance and reflectance regardless of the side of incidence (left or right). Fig. 3(a,c) shows that compared to an NBG, the bistability occurs at lower input intensities for the signals incident from the left of the grating and at higher intensity for signals incident from the right side of the grating. It is noted that the transmittances for the left and right incidence are different, $T_L \neq T_R$ as shown in Fig. 3(a,b), showing that the NPTBG does not satisfy Lorentzian reciprocity. This is due to the fact that the scattering matrix is no longer a complex symmetrical matrix $\bar{\bar{S}} \neq (\bar{\bar{S}})^t$. Furthermore, it is observed that at high intensity, both $\Gamma_L$ and $\Gamma_R$ are very low while transmittances are almost unity, implying bidirectional transparency (Fig. 4(c,d)).

When the NPTBG is operated with very low saturation intensity, e.g. $I_S = 65.2 \times 10^7$ Wm$^{-2}$, as taken from [21], and for the same input field intensity range the saturation factor varies in the range of $1.5 \times 10^{-6} < \mathbb{S} < 20 \times 10^{-6}$, it is observed that, regardless of the amount of gain and loss in the system, all results overlap with that of the NBG (dashed line on Fig. 3), i.e $T_L = T_R$ and $\Gamma_L = \Gamma_R$. This result, which is not shown separately in this paper, confirms that when gain and loss saturation intensity are very low, PT behavior is inhibited due to the negligible effective gain and loss.

We turn our attention now to the case of no gain/loss saturation, i.e. $\mathbb{S} = 1$. Fig. 4(a-d) shows that the NPTBG is bistable for both transmittance and reflectance regardless of the input signal incident (from left or right side). It is noted that in the absence of gain/loss saturation, $\mathbb{S} = 1$, both the width and on/off ratio of hysteresis reduce as the gain/loss in the grating is increased. Similarly as in Fig. 4(a,c), the bistability occurs at lower input intensities for the signals incident from the left of the grating compared to signals incident from the right. Of special interest is the U-operation ($|\alpha_0| = 1460.24$ cm$^{-1}$) at which the structure loses the hysteresis properties. Fig. 4(a,b) also shows that the transmittances for the left and right incidence are different, $T_L \neq T_R$, again showing that the NPTBG does not satisfy Lorentzian reciprocity. Fig. 5 shows the temporal response and frequency content of the transmitted signal for the left incident for input intensity $I = 2.6 \times 10^{14}$ Wm$^{-2}$, and shows the presence of longitudinal modes that fall within the gain/loss profile of the grating and are spaced at multiples of ~1.1 THz around the input signal frequency $f = 337.7$ THz. The rapid drop of the $T_L$ in Fig. 4(a) can thus be explained as a result of the transfer of energy to other frequencies.

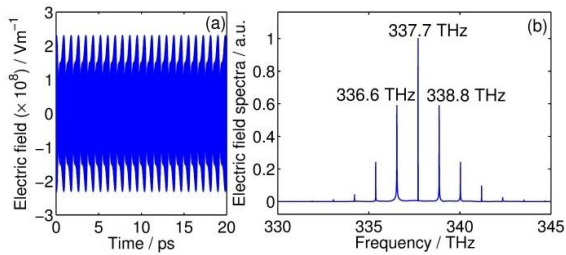

Fig. 5. (a) Time and (b) frequency response of the transmitted electric field of NPTBG at the U operation ($|\alpha_0| = 1460.24$ cm$^{-1}$) with input intensity $I = 2.6 \times 10^{14}$ Wm$^{-2}$ and the incident is from the left side of the grating.

In conclusion we have analyzed a nonlinear GaAs-based PT Bragg grating with a full dispersive and saturable gain and loss model and have demonstrated that levels of gain/loss saturation can have significant impact on PT behavior and should not be ignored. Low saturation intensity inhibits PT behavior and reduces the grating to a purely nonlinear grating, as we found it to be the case with the gain/loss saturation intensity reported in [21]. High saturation intensity enables an interplay of nonlinear and PT phenomena resulting in a reduction of the intensity levels at which bistability occurs. Finally, the presence of material dispersion limits the unidirectional invisibility to a narrow frequency range around the Bragg frequency and thus puts practical limits on cloaking applications.